\documentclass[sigconf]{acmart}

\usepackage{booktabs} 
\usepackage[english]{babel} 
\usepackage[utf8x]{inputenc} 
\usepackage[colorinlistoftodos]{todonotes} 

\setcopyright{rightsretained}

\begin{document}
\title[SPIRAL: Towards Better Participation of Older Adults in Software Development]{Guidelines Towards Better Participation of Older Adults in Software Development Processes using a new SPIRAL Method and Participatory Approach}



\author{Wies\l{}aw Kope\'{c}}
\affiliation{%
  \institution{Polish-Japanese Academy of Information Technology}
  \streetaddress{86 Koszykowa str.}
  \postcode{02-008}
  \city{Warsaw}
  \country{Poland}
}
\email{kopec@pja.edu.pl}

\author{Radoslaw Nielek}
\affiliation{%
  \institution{Polish-Japanese Academy of Information Technology}
  \streetaddress{86 Koszykowa str.}
  \postcode{02-008}
  \city{Warsaw}
  \country{Poland}
}
\email{nielek@pja.edu.pl}

\author{Adam Wierzbicki}
\affiliation{%
  \institution{Polish-Japanese Academy of Information Technology}
  \streetaddress{86 Koszykowa str.}
  \postcode{02-008}
  \city{Warsaw}
  \country{Poland}
}
\email{adamw@pja.edu.pl}

\renewcommand{\shortauthors}{W. Kope\'{c} et al.}

\begin{abstract}
This paper presents a new method of engaging older participants in the process of application and IT solutions development for older adults for emerging IT and tech startups. A new method called SPIRAL (Support for Participant Involvement in Rapid and Agile software development Labs) is proposed which adds both sustainability and flexibility to the development process with older adults. This method is based on the participatory approach and user empowerment of older adults with the aid of a bootstrapped Living Lab concept and it goes beyond well established user-centered and empathic design. SPIRAL provides strategies for direct involvement of older participants in the software development processes from the very early stage to support the agile approach with rapid prototyping, in particular in new and emerging startup environments with limited capabilities, including time, team and resources.

\end{abstract}




\copyrightyear{2018} 
\acmYear{2018} 
\setcopyright{acmcopyright}
\acmConference[CHASE'18]{CHASE'18:IEEE/ACM 11th International Workshop on Cooperative and Human Aspects of Software }{May 27, 2018}{Gothenburg, Sweden}
\acmPrice{15.00}
\acmDOI{10.1145/3195836.3195840}
\acmISBN{978-1-4503-5725-8/18/05}





\maketitle

\section{Introduction}
In light of the current socio-economic changes, demographic trends as well as the resulting need to develop tools and solutions for older adults we have decided to tackle the challenge of designing applications for, and with the older adults. Our solution is targeted at IT startup environments who operate with limited resources, but would like to follow industry best-practices or even go beyond them, by co-designing their solutions with the elderly end users. Therefore, below we outline our motivation towards creating a new method of participatory software development for older adults. 

\subsection{Demographic, social and economic trends}
An increase in life expectancy in developed countries, coupled with falling fertility rates brought about a demographic shift, which, according to United Nations estimate may mean that in 2050 older adults, defined as people aged 65+, will make up on average 20\% of the global population. At the same time, the number of people over 80 years old is expected to double. \cite{UNreport} Already in developed countries the share of older adults in the total population is sizable, with Japan leading the way at 27\% followed by Italy at 23\% and Germany at 21\%. \cite{worldbank} The challenges associated with this trend are not limited to the old-age dependency ratio,  the resulting lower tax revenues and supply shortages, but also include the need to shift the focus of key industries to meet the demands of changing markets, i.e. health care, education, but also consumer goods and technology.

The increasing number of older citizens equals the need to provide better tools and information for and about the needs of older adults to offset the effects of their deteriorating health and declining family support\cite{WHOreport} and at the same time to cater to their different expectations. This twofold pattern was dubbed as "Silver Economy", and currently it is estimated to be the third largest economy in the world, totaling about 7 trillion dollars per year. In the European Union the related government expenditure stands at 25\% of GDP and is projected to further increase. Owing to this, the initiatives related to silver economy are supported and actively developed by governments on many levels, with the Europe 2020 strategy calling for building opportunities for independent and active life for the elderly, who ought to voice their needs as one of the bigger consumer groups on the market. \cite{silverecon} 

\subsection{IT industry and software development trends}

New trends that emerged in the IT market over the recent years, such as the Internet of Things (IoT) and smart cities, are promising to present breakthroughs also in the field of active and assisted living for older adults. The software development industry making use of the agile methodology is growing, and the practice of rapid prototyping with it, especially in the context of mobile application market growth and the new startup economy. However, while more resources are being spent on R\&D the innovative outcomes grant lower returns on investment.\cite{knott2017rnd} 
In this context, insights from user participation are especially valuable. The agile startup environment is seeing a trend to develop methodologies with high potential of user involvement in the lean startup approach to design cycle.\citep{ries2011lean, blank2013four} There are multiple propositions based on user-centered and empathic design \cite{leonard1997spark} from a well-established design thinking method originating at Stanford University \cite{kelley2001art,kelley2005ten} to the recent Google Venture compact approach of design sprint.\cite{google2016sprint}
However, despite this massive effort, there is still a vast area of potential research in the field of effective methods and techniques for development of IT solutions, especially in the domain of software development for older adults. The existing solutions are either unsustainable, as they are resource intensive, or they lack flexibility to engage the users directly at every stage of the design process.

\subsection{Research problem}
According to Blank \cite{blank2013lean} the global advent of innovation economy is strongly driven by the expansion of startups. This expansion is happening in the context of the aforementioned socio-economic trends which arise from demographic and economic changes, as evidenced by Eurostat and OECD data and statistics \cite{silverecon}. The same trends are reflected in strategic planning by European Commission, in particular policies within Digital Single Market strategy, i.e. from boosting e-commerce, ICT innovation to Digital Inclusion for a better society, Aging Well with ICT and Startup Europe. Therefore it becomes more clear that there are increasingly more opportunities for emerging agile startups to create applications for older adults to tap into the rapidly growing silver economy.

On the other hand these new lean and emerging startups with limited resources have to face the problem of getting reliable knowledge about their end users. In effect they either work based on their best judgment or order expensive research and analyses. Moreover, most of them are small development teams of young people with little resources to spare. The average tech startup employee is below 30, and only 15\% are over 40 years old\cite{average_tech_startup}\cite{average_startup}. In light of these the alternative to big spending is to incorporate the practice of participatory design into their agile method and invite the potential users to co-design the solutions. 

However, older adults are a specific group, which may not be ready to join design activities without preparation. First, this group is hard to define, as it is not uniform - which makes treating them as a target group even more challenging. There are multiple approaches to defining the groups' boundaries and they include arbitrary age cutoff of 65+, the retirement age which may differ between states, but it may reflect economic and cultural differences or the time when people choose to leave employment permanently. For our purpose we use a combination of these three factors to control for outliers, such as early retirement or stay-at-home mothers. Thus,  we arrive at a group which shares some general characteristics.\cite{who_age}

Older adults, defined as such, are typically not very IT-literate and comfortable with technology, they lack confidence in their ICT skills and judgment, so they tend to accept the ideas of others above their own, when it comes to technology-related matters. While this is slowly changing, as progressively more tech-savvy people enter retirement, especially in less developed countries it is still common. On top of this young developers hold unconscious age-dependent biases about the older adults which prevent them from benefiting the fullest from their cooperation and insights. 

Therefore, at the intersection of these demographic and IT industry trends lies promising and largely untapped ground for proposing a sustainable, yet flexible, method of software development for older adults, which takes into account findings of original research related these barriers and challenges and the best practices of participatory design. This new method, called SPIRAL is what we propose and explore in this paper.
\\

The rest of the article is organized as follows. First we present previous works related to engaging older adults in context of human-centered approach and some software development concepts, processes, methods and frameworks that are promising in relation to older adults as participants.
This is followed by a brief description of trends in new participatory approach in software development for older adults. Next, we mention major challenges and limitations of presented methods. Finally we arrive at a detailed description of our new SPIRAL method and approaches with some practical examples from our research activities, together with guidelines and conclusions. Lastly, we explore future work potential.

\section{Related Work}

\subsection{Co-design concept}

\begin{figure}
\centering
\includegraphics[width=2in]{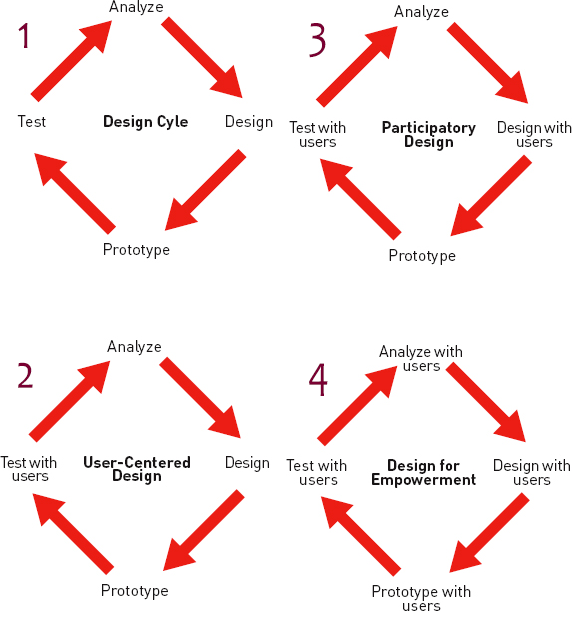}
\caption{Design process with end user engagement (Ladner, 2015)}
\label{fig:project_ladner}
\end{figure}

\begin{figure}
\centering
\includegraphics[width=3in]{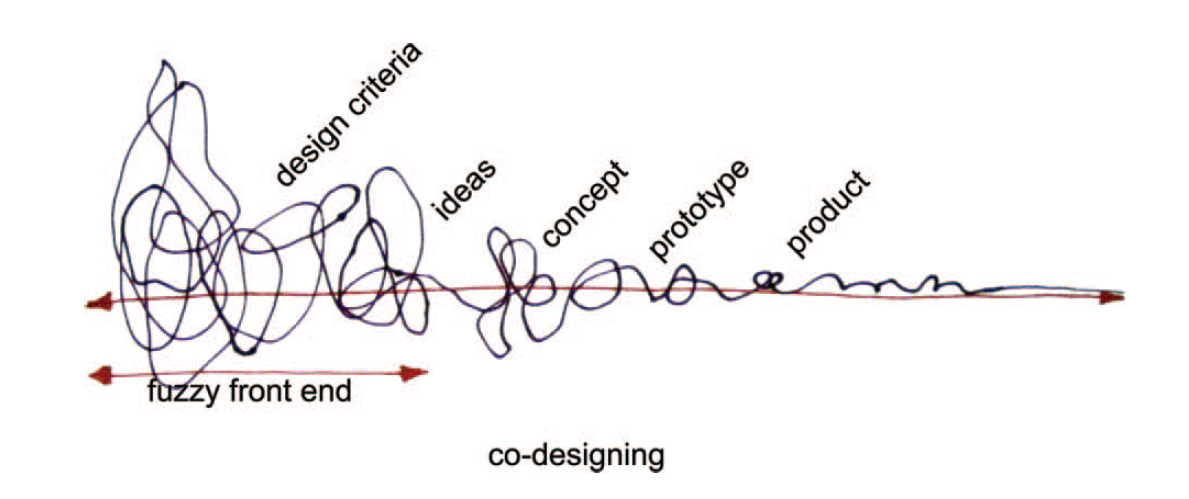}
\caption{Project development phases for co-design (Sanders \& Stappers, 2008)}
\label{fig:fuzzy}
\end{figure}

\begin{figure}
\centering
\includegraphics[height=200px]{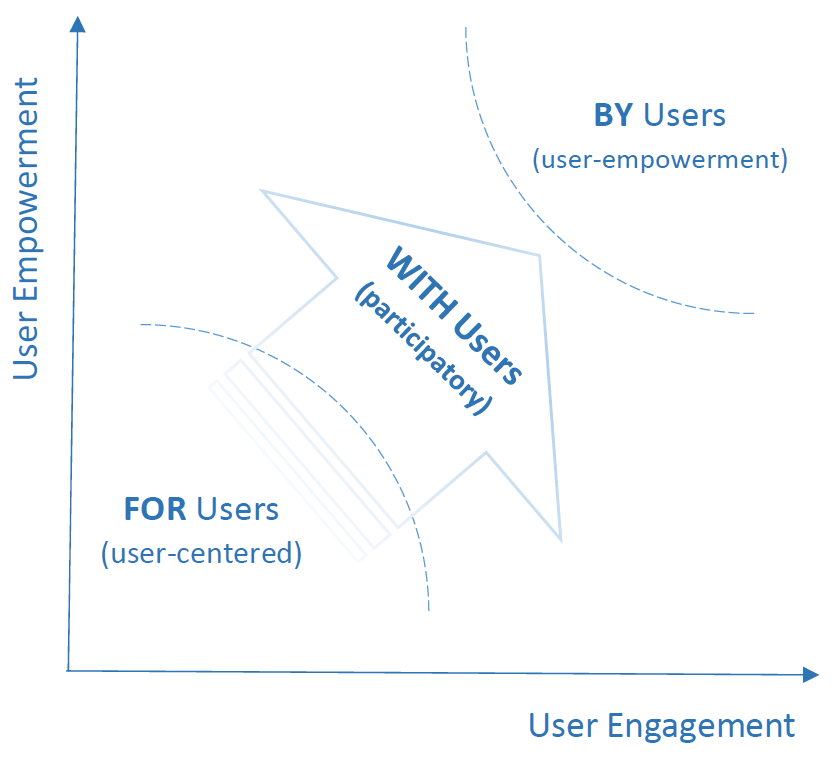}
\caption{User engagement and empowerment}
\label{fig:methods}
\end{figure}

Most of the current trends in software development which include the methods and techniques mentioned in the introductory section are closely related to human-centered approach with a participatory component. Moreover, the concept of human-centered approach is advocated for not only in IT solution development, but also in other diverse fields, such as education, health care or even architecture. \cite{szebeko2010co} Some of the related practices, like design sprint developed by Google Venture \cite{google2016sprint}
propose extensive participation of diverse stakeholders outside the core development team, but are mostly limited to various divisions from the company. Most of them are based on well-established user-centered approach and empathic design and in effect almost none of them stipulate direct engagement of real target users nor provide ready guidelines of how to enable such participation, in particular in relation do older adults. This is surprising, given that while the practice of user-centered design is well-established within human-centered approach, participatory design, or co-design, which goes further, to include them in the development process itself by Sanders\cite{sanders2002user,sanders2008co} is not as common, despite being a natural next step.
Sanders, further, argues that the whole design cycle can benefit from direct user involvement, from the earliest pre-design stage depicted on Fig \ref{fig:fuzzy}, which stands in contrast to Ladner who also advocates participatory design, but at the same excludes direct user involvement from the analysis and prototype stages. \cite{ladner2015design}  
Yet, it is thanks to Ladner that another term emerges, namely Design for Empowerment, which has users participate in each stage of the cycle: analysis, design, prototyping and testing in the name of greater accessibility and empowerment of user groups (Fig \ref{fig:project_ladner}). Therefore this continuity of approaches and methods of user participation in design and development process is often described as a three step perspective of user involvement and engagement: from user-centered design (for users), participatory processes (with users) to full involvement (by users) called by Ladner user for empowerment, illustrated on Fig. \ref{fig:methods}.

The concept of co-design may look straightforward, but its direct incorporation into existing methods, techniques and approaches of software development, especially solutions dedicated to older adults is uncommon.

\subsection{Older adults engagement}
 
Having in mind the concept of direct user participation in the co-design process described above one of the most vital problems is user engagement. It is especially prominent in the context of older adults, because of the ICT-related challenges and the unique nature of their empowerment and motivation.
Majority of older adults who reap the benefits of using ICT-related skills in their lives are motivated by the appreciation of their usefulness and see the use of technology to be necessary.\cite{aula_learning_2004}. As an age group who are exceptionally conscious of their needs, they mostly express interest in using ICT to achieve their personal goals. \cite{djoub_ict_2013}. So, most commonly they use technology to communicate with family and friends, manage administrative tasks (bills and bank account) and develop their hobbies. \cite{boulton2007ageing}, \cite{naumanen_practices_2008}. 

Already 1/3 of people over the age of 75 have physical or mental impairments, which limit them to some extent. So, as longevity increases the considerations related to accessibility come more into focus.
\cite{comcomm} Thus, a major social benefit from older adults involvement with ICT tasks comes from their staying mentally active, as it can delay the onset of age-related ailments. \cite{kotteritzsch2014adaptive}. Similar benefits come from volunteering, as it helps mitigate the limiting effects of aging related processes. \cite{lum2005effects} Yet, especially in the context of startup and silver economy it is important to mention after Lindsay and Davidson \cite{lindsay2012engaging,davidson2013participatory} that designing solutions for older adults should not be equaled with addressing the list of identified functional impairments but rather it ought to be considered as an opportunity to explore their needs and desires.

However, older adults lack confidence in their ICT-related skills which forms a strong barrier to their involvement in IT tasks in general, and in the development processes in particular. This effect can be mitigated by positive social intergenerational contact, \cite{kopec2017location} and the feeling of working for the greater good.
This is why there is a need to develop a co-design approach tailored for the older adults, which not only takes into account ways to produce tech insights through observation, but also facilitates positive interaction, social inclusion, physical and mental activity, and opens the older adults to the possibility of fulfilling their needs with the use of technology. 

\subsection{Design with and by older adults}
One of the better developed and independently tested\cite{davidson2013participatory} approaches to facilitate user participation is OASIS (Open architecture for Accessible Services Integration and Standardization) which consists of four crucial stages: 1) identification and recruitment of stakeholders, 2) creation of gapped video materials for the idea generation stage, 3) exploratory group meetings and 4) low fidelity prototyping.\cite{lindsay2012engaging} While it is reported to be useful for generating insights to keep costs low by creating better tailored solutions in the development process this approach requires extensive preparation and it does not necessarily create a more expert user group as a result, as instead of educating and empowering the stakeholders, it initially focuses on generating insights through indirect means of MadLibs-like videos.

\subsection{Living Labs approach}

The concept of Living Labs is one of the most recent and promising trends in supporting IT solution development processes. The term \textit{Living Lab} was coined by William Mitchell from MIT \cite{niitamo2006state}
and was used to refer to the real environment, like a home or a city, where routines and everyday life interactions of users and new technology can be observed and recorded. This experimental environment rooted in the philosophy of user-centered research fosters the process of designing new, useful and acceptable products and services. Thus, the users become co-producers and are at the center of the whole development process.\cite{ballon2005test,schumacher2007living} 

The unique insights of the users provide substantial business value as they mitigate risks inherent to the development of new solutions.\cite{pallot2010living} As living labs are long-term projects and direct involvement of the users is central to the participatory design approach there exists the need to maintain the interest of the stakeholder community in the LivingLab research and development network and offer them added value for their participation, which can take the form of trainings, workshops, games and access to new ICT solutions.

While the benefits of insights into product development from the LivingLab communities are clear, the resources needed to create and maintain a fully-fledged LivingLab, such as a "town in miniature" with its community are massive and pose a major general barrier to using this solution. Thus, there exists the need for the creation of a smaller "boostrap setup" which is less resource-intensive. Instead of a large initial investment and building a comprehensive LivingLab there is the opportunity to focus on enabling the key features of such stakeholder community - maintaining their motivation and building their competences as expert users, testers and eventually, co-designers while providing them with activities beneficial to their well-being, all outside of the LivingLab infrastructure.

\section{Major challenges and barriers}

The trend to involve the users directly in the software development process as active participants is clear and beneficial, so solutions such as Living Labs and the OASIS framework have a well-deserved place in the academia and the industry. However, their full implementation is associated with massive initial investments of resources, especially time and money, and a sizable effort to maintain (i.e. the Living Lab facilities and community engagement) and adjust (i.e. the single-use OASIS approach) them is required. 

This means they are not feasible in most small and emerging agile development environments and would cause a significant strain on startup teams resources. Moreover, both of these approaches limit the involvement of participants to only some stages of the development process, with OASIS being solely limited to the idea generation stage, and the LivingLabs placing emphasis on the more passive involvement, such as user observation and testing. 

Therefore, to mitigate the barriers related to the involvement of older adults as users and even co-creators we propose a different, more sustainable and truly participatory approach that focuses on the key issues the aforementioned solutions address by distilling the experience to a handful of major challenges:

\begin{tabular}{| p{2cm} | p{5cm} |}
	\hline
	\textbf{Challenge} & \textbf{Elaboration} \\
	\hline
    technology barrier and lack of ICT skills & low ICT literacy and familiarity with technology, lack of knowledge of existing solutions and understanding of basic ICT concepts \cite{kopec2017living, nielek2017wikipedia,lindsay2012engaging}\\
    \hline
   lack of self-confidence and limited social involvement & low self-confidence connected to the feeling of alienation from the increasingly technology-dependent society \cite{kopec2017location,lindsay2012engaging}\\
	\hline
   stereotypes among young startup teams & age-related stereotypes concerning the activities, interests and mental capacities of older adults \cite{kopec2017older,allport1979nature}\\
   	\hline
   improper level of criticism among older adults & bias for and eagerness to accept the solutions of others with low criticism \cite{orzeszek2017design,davidson2013participatory}\\
	\hline
\end{tabular}

These challenges point to the need to involve the users in even more steps of the development process, including the earliest ones and those usually reserved for industry professionals and to include them to a greater extent, as only this will address all of the problems uncovered in related research and counter the limitations of existing solutions. 

We opt for facilitating an almost organic emergence of a more expert and confident group of older adults who can take part in participatory design activities with young development teams, outside of the resource-intensive context of fully-fledged LivingLabs and beyond the constraints of the OASIS framework, which does not include direct interaction between older end users and development teams.

Below we propose a research-backed model of how to do this having in mind three key issues:  the necessity to establish a sustainable framework for long-term development environments, the need to offer flexible solutions to emerging startup teams, and the wish to empower older adults to overcome the challenges they face within the participatory design setup.

\section{SPIRAL approach}

Based on our LivingLab experience and research \cite{kopec2017living,kopec2017location,kopec2017older,orzeszek2017design,nielek2017wikipedia,nielek2017emotions} agile prototyping and participatory design concepts, we propose a 4-step model named SPIRAL (Support for Participant Involvement in Rapid and Agile software development Labs) which consists of:

\begin{itemize}
\item lowering technology barrier by increasing ICT skills;
\item enabling direct involvement with technology in everyday context and a positive emotional experience;
\item exposing participants to intergenerational interaction: facing stereotypes with contact theory;
\item empowering older participants by providing introductory step-by-step and hands-on experience trainings.
\end{itemize}

\begin{figure}
\centering
\includegraphics[height=200px]{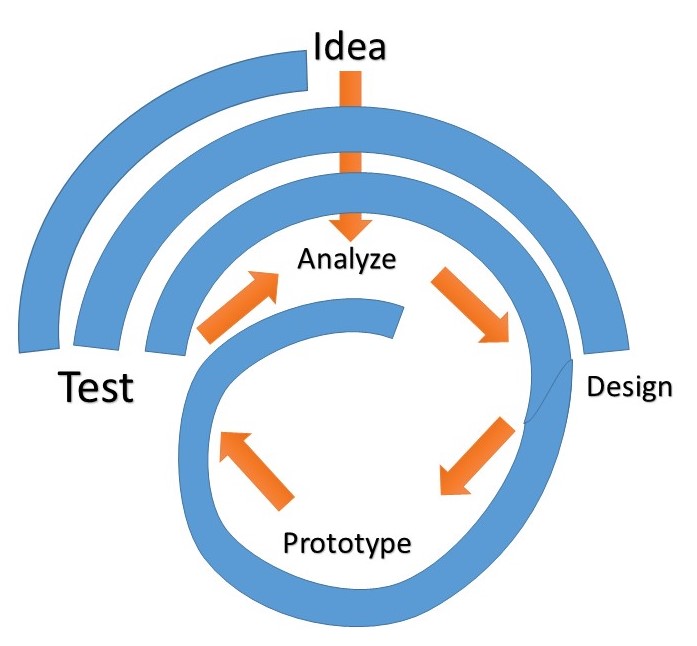}
\caption{SPIRAL user involvement scheme}
\label{fig:spiral_scheme}
\end{figure}
The spiral scheme illustrates the stages of user involvement in the development process, based on the competences of the users, which they grow in the course of their participation in design. The least advanced users may participate in the idea generation and testing phases, more advanced can join the design and analysis stages, whereas the most advanced can take part in the full development cycle as experts. As the competences of the users grow, they can join in more of the stages, not only contributing more and different insights into the process, but also engaging with ICT on their own, which promotes their social inclusion in technology dependent tasks.

Below we present each of the steps of the SPIRAL method with the description of basic roles and rules that should be helpful in the process of involving older adults in its various stages. While the SPIRAL method is our proposal, it consists of familiar elements, whose effectiveness is verified on their own - and its innovative proposition relies on the context of their use and their interplay relying on logical progression and continuity. Following each step in order should further the establishment of a sustainable lean LivingLab community, however having in mind the dynamics of agile processes and limited resources in lean startup approach we propose some shortcuts within the SPIRAL model.

\subsection{Lowering ICT barriers}
This is the first of the two preparatory steps.

\textbf{Purpose}: This step is devoted to lowering the technology barrier by addressing lack of ICT skills among older adults and providing the foundation towards sustainable community and crowdsourcing tasks.

\textbf{Basic roles}: senior (older adults), instructor and facilitator (instructor assistant), author and researcher (course and tasks creation), 

\textbf{Prerequisites or characteristic}: senior (none or low ICT literacy), instructor (experience in adult education), facilitator (preferred older adult or young)

\textbf{Tools}: e-platform with pre-made courses for blended learning, basic applications, simple pre-made crowdsourcing tasks. Paper and online tests (quizzes).

\textbf{Techniques and activities}: traditional computer courses and workshops, introductory courses for the e-platform, blended courses.

\textbf{Rules}: The taught basic ICT notions should offer some practical tools, they can use in their daily lives, such as e-mails, Skype or Facebook.

\textbf{Next step}: 2 (direct involvement with technology)

\textbf{Recommendations and notes}: Blended courses are more effective than self-paced ones. The role of the facilitator (instructor assistant) is very important,
and it is worth to mention that the interaction with the instructor might be perceived as one of the most important aspects by participants.\cite{rhode2009interaction} 
Quizzes for edutainment purposes (knowledge check and entertainment) are effective. Pre-made crowdsourcing tasks and surveys to prepare for better involvement in further steps and software development cycle can be given. This step can be omitted and older participants can start with second step, however in order to start the process of building a sustainable community it is recommended to go through this step. On the other hand swapping it with an extended second step could be effective, especially with mobile devices and applications as they are easier to explore.

\subsection{Direct involvement with technology} 

This is the second of the two preparatory steps.

\textbf{Purpose}: This step is devoted to direct involvement with technology usage in everyday context and a positive emotional experience. It could also bring some intergenerational interaction if younger participants are involved. In case of presence of a development team representative it could lead into some ice-breaking conditions, which could be fruitful in further steps focused more on the development process.

\textbf{Basic roles}: senior (older adults), facilitator (senior or junior)

\textbf{Prerequisites or characteristic}: senior (limited to moderate ICT literacy), facilitator (tech-savvy)

\textbf{Tools}: mobile devices and application

\textbf{Techniques and activities}: learning by doing, engaging activities, complex and free-flow tasks (complex crowdsourcing tasks), i.e. location-based game (exploration of IT tools), Wikipedia editing (content co-creation).

\textbf{Rules}: Facilitator can not take control of the device. Older adult should be the one that uses and explores the device and the application in order to break the barriers of using the technology and fears of breaking the device or the application.

\textbf{Next step}: 3 (direct interaction with developers).

\textbf{Recommendations and notes}: The facilitator can be either an older adult or a young person. Young person more advisable in order to prepare the older adult for further steps and better interaction with younger participants. Possible and advisable participation of software development team representatives for ice-breaking and facilitating direct interaction and cooperation in further steps.

\subsection{Intergenerational interaction with developers}

This is the first of the two software and interface design steps.

\textbf{Purpose}: This step is devoted to intergenerational interaction with software development team and facing stereotypes with the use of contact theory, which improves relations between different groups. This is a simulation of a whole design process in miniature, thus the form of the hackathon is most suitable. Within it, the involvement of older adults is needed at the beginning of the process with optional continuation and final presentation at the results of the work. The intention of this step is to provide an opportunity for immersion of older adults in the development process and facing stereotypes from both sides of the process.

\textbf{Basic roles}: senior team members (older adults) and junior team members (young developers and designers)

\textbf{Prerequisites or characteristic}: for seniors it is recommended to take part in the activities from the previous steps (1-2) (or to invite entry-level seniors just for the testing phase purposes).

\textbf{Tools}: hackathon or other live team cooperation environment with full design cycle from idea development to testing plus introductory presentation of themes, rules and prizes.

\textbf{Techniques}: direct cooperation, teamwork, intergenerational cooperation

\textbf{Rules}: Each participant should have equal right of voice. Participation is voluntary. Each member of the team can take a break at any time, but should remain accessible to other team members (either by phone or e-mail).

\textbf{Next step}: 4 (user empowerment)

\textbf{Recommendations and notes}: Hackathon team should consist of four to six people: two-three young developers and one-two designers with participation of two older adults. It should be organized as a competition between teams. Hackathon can be normal length (24-hour) or shorter, but it should start and end at the time feasible for older participants. Older participants do not need to be present throughout the full process. The most important are: the first part (idea development and analysis) alongside with the last part (testing the outcomes and results of development team). It is advisable to inform the parties that after the introductory phase they can cooperate as they see fit, in full and up to, or from any stage of the development process.

\subsection{Participant empowerment}

This is a second of the two design steps.

\textbf{Purpose}: This step is devoted to empowerment of older participants alongside with enriching young developers and designers skills of cooperation with older participants in co-design mode.

\textbf{Basic roles}: senior (older adults), junior (designer or developer)

\textbf{Prerequisites or characteristic}: ICT literate senior with experience in direct cooperation (recommended previous steps) and a young developer with similar experience (recommended previous step).

\textbf{Tools}: Co-design contest in limited time for enabling cooperation for the whole process, not only the idea development stage. User empowerment workshops for older adults introducing required part of hands-on experience trainings starting from UI towards software development.

\textbf{Techniques}: Direct intergenerational cooperation, pair design and prototyping.

\textbf{Rules}: Senior and junior have equal rights to take part and contribute in the process. Unlike the previous step the contest can omit the idea generation stage and be limited to analysis, design, prototyping and testing initially limited to the layer of user interface, i.e. based on work on live mock-ups.

\textbf{Next step}: Further training in various aspects of interface and software design and cooperation work.

\textbf{Recommendations and notes}: Starting from individual one-one short time contest that is intended to enable cooperation throughout the whole process, and not focused predominantly on the idea development and testing stage like in previous step. This can lead to more in-depth SPIRAL process, from interface and live mock-up development towards more sophisticated and sustainable user involvement in software development process.

\section{SPIRAL in action}
Based on our previous research and papers we can provide some examples that proved effectiveness of particular steps from the proposed SPIRAL model.

\textbf{Lowering ICT barriers}. In cooperation with the City of Warsaw we have prepared and deployed the dedicated e-Senior platform for engaging older adults into sustainable bootstrapped LivingLab environment that attracted several hundred senior citizens.
Further information about our Living Lab and activities with older adults was presented on Web Intelligence conference. \cite{kopec2017living}

\begin{figure}
\centering
\includegraphics[height=130px]{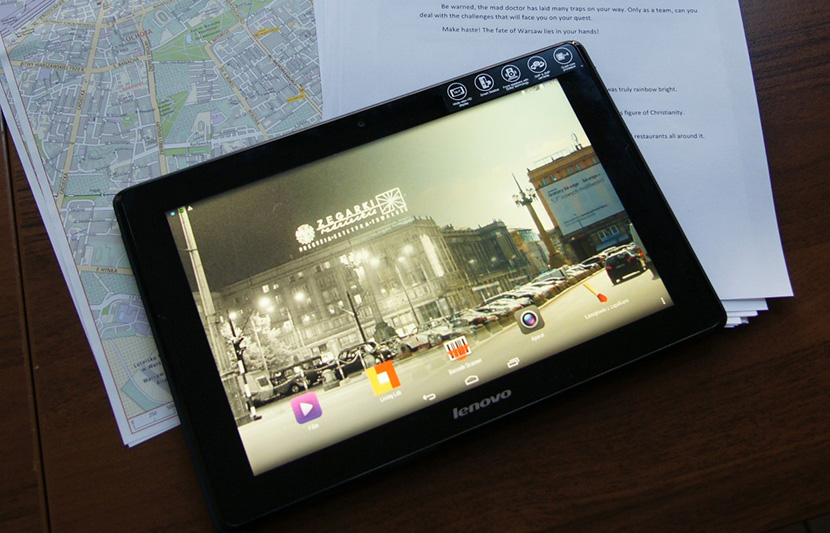}
\caption{Location-based game mobile application}
\label{fig:game_app}
\end{figure}
\begin{figure}
\centering
\includegraphics[height=130px]{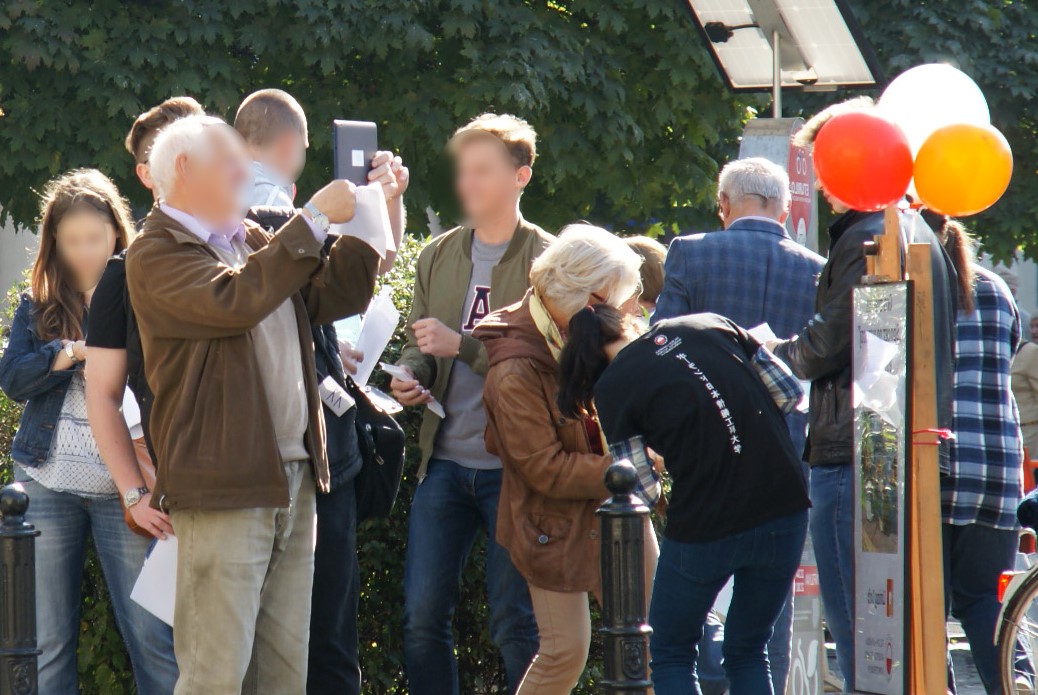}
\caption{Direct involvement -- an activity during the location-based game}
\label{fig:game_panoramic}
\end{figure}

\textbf{Direct involvement with technology}. An example of an activity which positively introduced older adults into the mobile technology usage was a location-based game \cite{kopec2017location}. Within it, we took our users on an intergenerational adventure in which they uncovered a mystery related to the history of Warsaw with the use of tablets (Fig. \ref{fig:game_app}, \ref{fig:game_panoramic}). A quick overview of the concept is depicted in the online movie\footnote{$https://youtu.be/htNieG0FwfY$}. Additionally, we also engaged them with increasingly complex crowdsourcing tasks.

\begin{figure}
\centering
\includegraphics[height=130px]{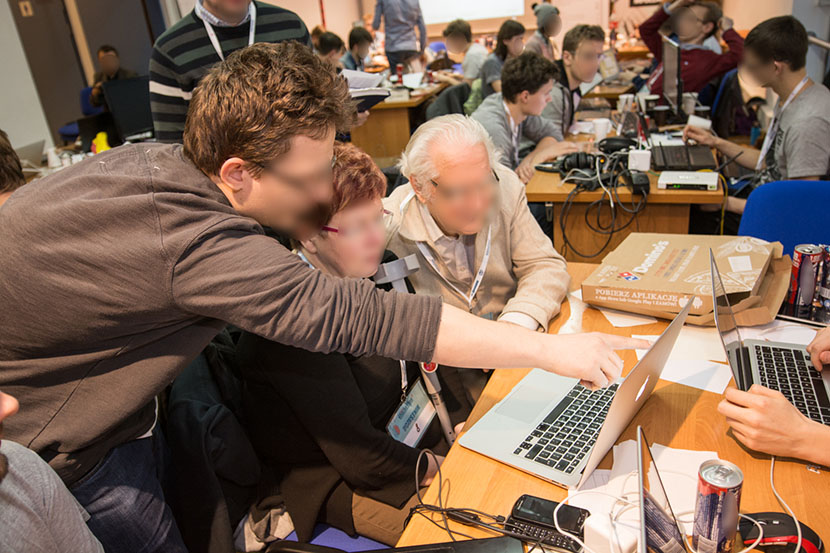}
\caption{Direct intergenerational cooperation during the hackathon}
\label{fig:hackathon_screen}
\end{figure}

\textbf{Intergenerational interaction with developers}. Co-design in teams, e.g. hackathon  proved that this kind of interaction can benefit both sides of the process (Fig. \ref{fig:hackathon_screen}). Brief overview of the hackathon for over a hundred participants is available online\footnote{$https://youtu.be/13aEBkrxe20$}. The method and research were thoroughly described in \textit{Empirical Software Engineering} \cite{kopec2017older} and were also a subject of corresponding ICSE journal-first session.\cite{kopec2017older_icse}

\begin{figure}
\centering
\includegraphics[height=130px]{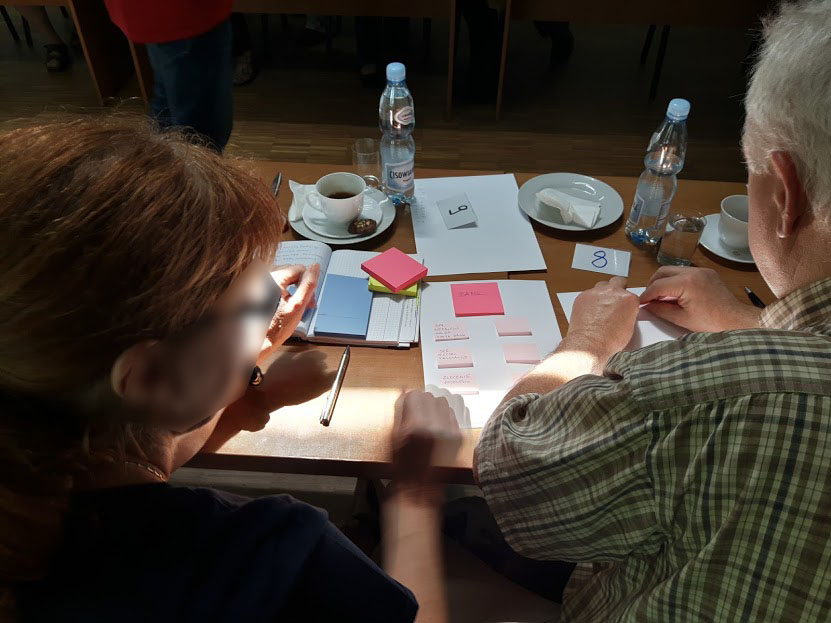}
\caption{User empowerment workshops -- paper prototyping}
\label{fig:design_paper}
\end{figure}

\begin{figure}
\centering
\includegraphics[height=130px]{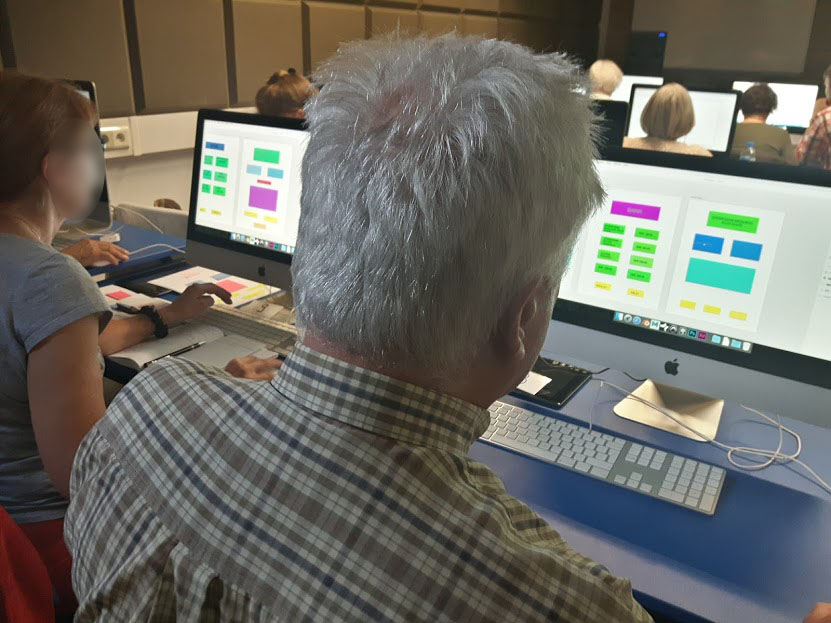}
\caption{User empowerment workshops -- digital prototyping}
\label{fig:design_digital}
\end{figure}

\textbf{Participants empowerment}: We have successfully conducted a series of user empowering workshops on the full cycle of designing mobile applications interfaces, including both paper (Fig. \ref{fig:design_paper}) and digital prototypes (Fig. \ref{fig:design_digital}) workshops, alongside with the final live mock-up development during intergenerational pair co-design contest. 
A brief overview of the whole process is available online, and while research is still in progress, some interesting findings were  published\cite{orzeszek2017design}. We also have some experience in other activities with significant empowerment potential like Wikipedia content co-creation depicted in a promising exploratory study \cite{nielek2017wikipedia}. Another example of an engaging activity that empowered older adults were workshops dedicated to prototyping a game for and with older adults. 
The work is in progress but interesting insights have been obtained and user engagement studies will be continued.

\section{Conclusions and further work}

The proposed SPIRAL method addresses ways to counter the barriers to older adults' participation in the work of startup development teams related to their ICT-literacy, empowerment and self-confidence. It is especially helpful for intergenerational cooperation - to encourage older adults to evaluate the work of younger developers, and to challenge  young developers' stereotypes concerning older adults with contact theory-inspired tools. In this paper we share our SPIRAL blueprint with the community not only to encourage its exploration by our fellow researchers and developers, but also to lay out plans to explore its limits and develop it further so that it can be used not only for more effective and empowering work with older adults, but also for enabling cooperation with other vulnerable groups within our society. While SPIRAL was developed with older adults in mind we believe that its basic design assumptions including the identification of key barriers to user participation and addressing them by a structured approach could be used with other vulnerable and underrepresented target groups.

Involving all vulnerable groups already at the very first stages of the development process is especially important in the youth-saturated startup scene as developers often hold unconscious biases. It is a challenge to explore existing needs, capacities and preferences of the "others" who they do not have much contact with, including older adults with whom they may interact more often, albeit in different contexts. Thus, early, varied and thorough engagement of users, already at the idea generation stage and up to testing, prevents stereotypes from surfacing and improves the functionality and accessibility of the conceptualized solutions. 

Additionally, when it comes to older adults, their deep empowerment which comes from prior training in ICT-related skills and experience co-designing applications in intergenerational teams makes them more likely to further pursue their interests using ICT as a tool. This, in turn, enables them to lead more active and independent lives. In this process they become natural expert users without external training, and multiple opportunities open up to them, as they may take part in crowd-sourcing and citizen science initiatives. This transformation also enables them to engage in deeper and more fruitful cooperation in software development processes in particular in current recent and vivid trends like IoT and smart cities in the context of meeting their needs within the silver economy.

Therefore, in the future studies we intend to research ways in which we can further mitigate the barriers to participation of older adults in ICT-related tasks, through earlier and deeper involvement in their design, as well as study the potential of new interfaces (i.e. SmartTVs, Chatbots, VR) for engaging older adults with technology related tasks of high social value, such as crowdsourcing, while further building up on their innate strengths of the wealth of experience and proficiency in their native language. In the development of these solutions we intend to benefit from the co-design insights coming from the SPIRAL-empowered users engaging in participatory design. Moreover, we would like to explore more ways of empowering and involving other vulnerable groups.

\section{Acknowledgments}
This project has received funding from the European Union's Horizon 2020 research and innovation programme under the Marie Sklodowska-Curie grant agreement No 690962 and from the Polish National Science Centre grant 2015/19/B/ST6/03179.

\bibliographystyle{ACM-Reference-Format}
\bibliography{sigproc} 


\begin{thebibliography}{00}


\ifx \showCODEN    \undefined \def \showCODEN     #1{\unskip}     \fi
\ifx \showDOI      \undefined \def \showDOI       #1{#1}\fi
\ifx \showISBNx    \undefined \def \showISBNx     #1{\unskip}     \fi
\ifx \showISBNxiii \undefined \def \showISBNxiii  #1{\unskip}     \fi
\ifx \showISSN     \undefined \def \showISSN      #1{\unskip}     \fi
\ifx \showLCCN     \undefined \def \showLCCN      #1{\unskip}     \fi
\ifx \shownote     \undefined \def \shownote      #1{#1}          \fi
\ifx \showarticletitle \undefined \def \showarticletitle #1{#1}   \fi
\ifx \showURL      \undefined \def \showURL       {\relax}        \fi
\providecommand\bibfield[2]{#2}
\providecommand\bibinfo[2]{#2}
\providecommand\natexlab[1]{#1}
\providecommand\showeprint[2][]{arXiv:#2}

\bibitem[\protect\citeauthoryear{??}{ave}{2002}]%
        {average_tech_startup}
 \bibinfo{year}{2002}\natexlab{}.
\newblock \bibinfo{title}{Here's What The Average Tech Startup Looks Like}.
\newblock   (\bibinfo{year}{2002}).
\newblock
\showURL{%
\url{http://www.businessinsider.com/heres-what-the-average-tech-startup-looks-like-2011-12}}


\bibitem[\protect\citeauthoryear{??}{who}{2002}]%
        {who_age}
 \bibinfo{year}{2002}\natexlab{}.
\newblock \bibinfo{title}{Proposed working definition of an older person in
  Africa for the MDS Project}.
\newblock   (\bibinfo{year}{2002}).
\newblock
\showURL{%
\url{http://www.who.int/healthinfo/survey/ageingdefnolder/en/}}


\bibitem[\protect\citeauthoryear{??}{com}{2010}]%
        {comcomm}
 \bibinfo{year}{2010}\natexlab{}.
\newblock \bibinfo{booktitle}{{\em Communication from the Commission to the
  European Parliament, the Council, the European Economic and Social Commitee
  and the Committee of the Regions}}.
\newblock \bibinfo{type}{{T}echnical {R}eport}. \bibinfo{institution}{European
  Commission}.
\newblock


\bibitem[\protect\citeauthoryear{??}{ave}{2014}]%
        {average_startup}
 \bibinfo{year}{2014}\natexlab{}.
\newblock \bibinfo{title}{Statistically Speaking, What Does the Average Startup
  Look Like?}
\newblock   (\bibinfo{year}{2014}).
\newblock
\showURL{%
\url{https://www.theatlantic.com/business/archive/2014/12/statistically-speaking-what-does-the-average-startup-look-like/384019/}}


\bibitem[\protect\citeauthoryear{??}{sil}{2015}]%
        {silverecon}
 \bibinfo{year}{2015}\natexlab{}.
\newblock \bibinfo{booktitle}{{\em Growing the European Silver Economy}}.
\newblock \bibinfo{type}{{T}echnical {R}eport}. \bibinfo{institution}{European
  Com.}
\newblock


\bibitem[\protect\citeauthoryear{??}{UNr}{2015}]%
        {UNreport}
 \bibinfo{year}{2015}\natexlab{}.
\newblock \bibinfo{booktitle}{{\em World Population Aging 2015}}.
\newblock \bibinfo{type}{{T}echnical {R}eport}. \bibinfo{institution}{United
  Nations Department of Economic and Social Affairs}. \bibinfo{pages}{9--39}
  pages.
\newblock


\bibitem[\protect\citeauthoryear{??}{WHO}{2016}]%
        {WHOreport}
 \bibinfo{year}{2016}\natexlab{}.
\newblock \bibinfo{booktitle}{{\em Global health and ageing}}.
\newblock \bibinfo{type}{{T}echnical {R}eport}. \bibinfo{institution}{World
  Health Organization}.
\newblock


\bibitem[\protect\citeauthoryear{??}{wor}{2016}]%
        {worldbank}
 \bibinfo{year}{2016}\natexlab{}.
\newblock \bibinfo{title}{Population ages 65 and above}.
\newblock   (\bibinfo{year}{2016}).
\newblock
\showURL{%
\url{https://data.worldbank.org/}}


\bibitem[\protect\citeauthoryear{Allport}{Allport}{1979}]%
        {allport1979nature}
\bibfield{author}{\bibinfo{person}{Gordon~Willard Allport}.}
  \bibinfo{year}{1979}\natexlab{}.
\newblock \bibinfo{booktitle}{{\em The nature of prejudice}}.
\newblock \bibinfo{publisher}{Basic books}.
\newblock


\bibitem[\protect\citeauthoryear{Aula}{Aula}{2004}]%
        {aula_learning_2004}
\bibfield{author}{\bibinfo{person}{Anne Aula}.}
  \bibinfo{year}{2004}\natexlab{}.
\newblock \showarticletitle{Learning to use computers at a later age}.
\newblock In \bibinfo{booktitle}{{\em {HCI} and the {Older} {Population}}}.
  \bibinfo{publisher}{University of Glasgow}, \bibinfo{address}{Leeds, UK},
  \bibinfo{pages}{3--5}.
\newblock


\bibitem[\protect\citeauthoryear{Ballon, Pierson, and Delaere}{Ballon
  et~al\mbox{.}}{2005}]%
        {ballon2005test}
\bibfield{author}{\bibinfo{person}{Pieter Ballon}, \bibinfo{person}{Jo
  Pierson}, {and} \bibinfo{person}{Simon Delaere}.}
  \bibinfo{year}{2005}\natexlab{}.
\newblock \showarticletitle{Test and experimentation platforms for broadband
  innovation: Examining European practice}.
\newblock  (\bibinfo{year}{2005}).
\newblock


\bibitem[\protect\citeauthoryear{Blank}{Blank}{2013a}]%
        {blank2013four}
\bibfield{author}{\bibinfo{person}{Steve Blank}.}
  \bibinfo{year}{2013}\natexlab{a}.
\newblock \bibinfo{booktitle}{{\em The four steps to the epiphany}}.
\newblock \bibinfo{publisher}{BookBaby}.
\newblock


\bibitem[\protect\citeauthoryear{Blank}{Blank}{2013b}]%
        {blank2013lean}
\bibfield{author}{\bibinfo{person}{Steve Blank}.}
  \bibinfo{year}{2013}\natexlab{b}.
\newblock \showarticletitle{Why the lean start-up changes everything}.
\newblock \bibinfo{journal}{{\em Harvard business review\/}}
  \bibinfo{volume}{91}, \bibinfo{number}{5} (\bibinfo{year}{2013}),
  \bibinfo{pages}{63--72}.
\newblock


\bibitem[\protect\citeauthoryear{Boulton-Lewis, Buys, Lovie-Kitchin, Barnett,
  and David}{Boulton-Lewis et~al\mbox{.}}{2007}]%
        {boulton2007ageing}
\bibfield{author}{\bibinfo{person}{Gillian~M Boulton-Lewis},
  \bibinfo{person}{Laurie Buys}, \bibinfo{person}{Jan Lovie-Kitchin},
  \bibinfo{person}{Karen Barnett}, {and} \bibinfo{person}{L~Nikki David}.}
  \bibinfo{year}{2007}\natexlab{}.
\newblock \showarticletitle{Ageing, learning, and computer technology in
  Australia}.
\newblock \bibinfo{journal}{{\em Educational Gerontology\/}}
  \bibinfo{volume}{33}, \bibinfo{number}{3} (\bibinfo{year}{2007}),
  \bibinfo{pages}{253--270}.
\newblock


\bibitem[\protect\citeauthoryear{Davidson and Jensen}{Davidson and
  Jensen}{2013}]%
        {davidson2013participatory}
\bibfield{author}{\bibinfo{person}{Jennifer~L Davidson} {and}
  \bibinfo{person}{Carlos Jensen}.} \bibinfo{year}{2013}\natexlab{}.
\newblock \showarticletitle{Participatory design with older adults: an analysis
  of creativity in the design of mobile healthcare applications}. In
  \bibinfo{booktitle}{{\em Proceedings of the 9th ACM Conference on Creativity
  \& Cognition}}. ACM, \bibinfo{pages}{114--123}.
\newblock


\bibitem[\protect\citeauthoryear{Djoub}{Djoub}{2013}]%
        {djoub_ict_2013}
\bibfield{author}{\bibinfo{person}{Zineb Djoub}.}
  \bibinfo{year}{2013}\natexlab{}.
\newblock \showarticletitle{{ICT} education and motivating elderly people}. In
  \bibinfo{booktitle}{{\em Ariadna}}, Vol.~\bibinfo{volume}{1}.
  \bibinfo{pages}{88--92}.
\newblock


\bibitem[\protect\citeauthoryear{Kelley and Littman}{Kelley and
  Littman}{2001}]%
        {kelley2001art}
\bibfield{author}{\bibinfo{person}{Tom Kelley} {and} \bibinfo{person}{Jonathan
  Littman}.} \bibinfo{year}{2001}\natexlab{}.
\newblock \bibinfo{booktitle}{{\em The art of innovation: Lessons in creativity
  from IDEO, America's leading design firm}}.
\newblock \bibinfo{publisher}{Broadway Business}.
\newblock


\bibitem[\protect\citeauthoryear{Kelley and Littman}{Kelley and
  Littman}{2005}]%
        {kelley2005ten}
\bibfield{author}{\bibinfo{person}{Tom Kelley} {and} \bibinfo{person}{Jonathan
  Littman}.} \bibinfo{year}{2005}\natexlab{}.
\newblock \bibinfo{booktitle}{{\em The Ten Faces of Innovation: IDEO's
  Strategies for Beating the Devil's Advocate \& Driving Creativity Throughout
  Your Organization}}.
\newblock \bibinfo{publisher}{Broadway Business}.
\newblock


\bibitem[\protect\citeauthoryear{Knapp, Zeratsky, and Kowitz}{Knapp
  et~al\mbox{.}}{2016}]%
        {google2016sprint}
\bibfield{author}{\bibinfo{person}{Jake Knapp}, \bibinfo{person}{John
  Zeratsky}, {and} \bibinfo{person}{Braden Kowitz}.}
  \bibinfo{year}{2016}\natexlab{}.
\newblock \bibinfo{booktitle}{{\em Sprint: How to solve big problems and test
  new ideas in just five days}}.
\newblock \bibinfo{publisher}{Simon and Schuster}.
\newblock


\bibitem[\protect\citeauthoryear{Knott}{Knott}{2017}]%
        {knott2017rnd}
\bibfield{author}{\bibinfo{person}{Anne~Marie Knott}.}
  \bibinfo{year}{2017}\natexlab{}.
\newblock \showarticletitle{Is R\&D Getting Harder, or Are Companies Just
  Getting Worse At It?}
\newblock \bibinfo{journal}{{\em Harvard Business Review\/}}
  (\bibinfo{year}{2017}).
\newblock


\bibitem[\protect\citeauthoryear{Kope{\'c}, Abramczuk, Balcerzak, Ju{\'z}win,
  Gniadzik, Kowalik, and Nielek}{Kope{\'c} et~al\mbox{.}}{2017a}]%
        {kopec2017location}
\bibfield{author}{\bibinfo{person}{Wies{\l}aw Kope{\'c}},
  \bibinfo{person}{Katarzyna Abramczuk}, \bibinfo{person}{Bart{\l}omiej
  Balcerzak}, \bibinfo{person}{Marta Ju{\'z}win}, \bibinfo{person}{Katarzyna
  Gniadzik}, \bibinfo{person}{Grzegorz Kowalik}, {and}
  \bibinfo{person}{Rados{\l}aw Nielek}.} \bibinfo{year}{2017}\natexlab{a}.
\newblock \showarticletitle{A Location-Based Game for Two Generations: Teaching
  Mobile Technology to the Elderly with the Support of Young Volunteers}.
\newblock In \bibinfo{booktitle}{{\em eHealth 360}}.
  \bibinfo{publisher}{Springer}, \bibinfo{pages}{84--91}.
\newblock


\bibitem[\protect\citeauthoryear{Kope{\'c}, Balcerzak, Nielek, Kowalik,
  Wierzbicki, and Casati}{Kope{\'c} et~al\mbox{.}}{2017b}]%
        {kopec2017older}
\bibfield{author}{\bibinfo{person}{Wies{\l}aw Kope{\'c}},
  \bibinfo{person}{Bart{\l}omiej Balcerzak}, \bibinfo{person}{Rados{\l}aw
  Nielek}, \bibinfo{person}{Grzegorz Kowalik}, \bibinfo{person}{Adam
  Wierzbicki}, {and} \bibinfo{person}{Fabio Casati}.}
  \bibinfo{year}{2017}\natexlab{b}.
\newblock \showarticletitle{Older adults and hackathons: a qualitative study}.
\newblock \bibinfo{journal}{{\em Empirical Software Engineering\/}}
  (\bibinfo{year}{2017}), \bibinfo{pages}{1--36}.
\newblock
\showDOI{%
\url{https://doi.org/10.1007/s10664-017-9565-6}}


\bibitem[\protect\citeauthoryear{Kope{\'c}, Balcerzak, Nielek, Kowalik,
  Wierzbicki, and Casati}{Kope{\'c} et~al\mbox{.}}{2018}]%
        {kopec2017older_icse}
\bibfield{author}{\bibinfo{person}{Wies{\l}aw Kope{\'c}},
  \bibinfo{person}{Bart{\l}omiej Balcerzak}, \bibinfo{person}{Rados{\l}aw
  Nielek}, \bibinfo{person}{Grzegorz Kowalik}, \bibinfo{person}{Adam
  Wierzbicki}, {and} \bibinfo{person}{Fabio Casati}.}
  \bibinfo{year}{2018}\natexlab{}.
\newblock \showarticletitle{Older adults and hackathons: a qualitative study}.
  In \bibinfo{booktitle}{{\em Proceedings of the 40th International Conference
  on Software Engineering}} {\em (\bibinfo{series}{ICSE '18})}.
  \bibinfo{publisher}{ACM}, \bibinfo{address}{New York, NY, USA}.
\newblock
\showDOI{%
\url{https://doi.org/10.1145/3180155.3182547}}


\bibitem[\protect\citeauthoryear{Kope\'{c}, Skorupska, Jaskulska, Abramczuk,
  Nielek, and Wierzbicki}{Kope\'{c} et~al\mbox{.}}{2017}]%
        {kopec2017living}
\bibfield{author}{\bibinfo{person}{Wies{\l}aw Kope\'{c}},
  \bibinfo{person}{Kinga Skorupska}, \bibinfo{person}{Anna Jaskulska},
  \bibinfo{person}{Katarzyna Abramczuk}, \bibinfo{person}{Radoslaw Nielek},
  {and} \bibinfo{person}{Adam Wierzbicki}.} \bibinfo{year}{2017}\natexlab{}.
\newblock \showarticletitle{LivingLab PJAIT: Towards Better Urban Participation
  of Seniors}. In \bibinfo{booktitle}{{\em Proceedings of the International
  Conference on Web Intelligence}} {\em (\bibinfo{series}{WI '17})}.
  \bibinfo{publisher}{ACM}, \bibinfo{address}{New York, NY, USA},
  \bibinfo{pages}{1085--1092}.
\newblock
\showISBNx{978-1-4503-4951-2}
\showDOI{%
\url{https://doi.org/10.1145/3106426.3109040}}


\bibitem[\protect\citeauthoryear{K{\"o}tteritzsch, Koch, and
  Lem{\^a}n}{K{\"o}tteritzsch et~al\mbox{.}}{2014}]%
        {kotteritzsch2014adaptive}
\bibfield{author}{\bibinfo{person}{Anna K{\"o}tteritzsch},
  \bibinfo{person}{Michael Koch}, {and} \bibinfo{person}{Fritjof Lem{\^a}n}.}
  \bibinfo{year}{2014}\natexlab{}.
\newblock \showarticletitle{Adaptive training for older adults based on dynamic
  diagnosis of mild cognitive impairments and dementia}. In
  \bibinfo{booktitle}{{\em International Workshop on Ambient Assisted Living}}.
  Springer, \bibinfo{pages}{364--368}.
\newblock


\bibitem[\protect\citeauthoryear{Ladner}{Ladner}{2015}]%
        {ladner2015design}
\bibfield{author}{\bibinfo{person}{Richard~E Ladner}.}
  \bibinfo{year}{2015}\natexlab{}.
\newblock \showarticletitle{Design for user empowerment}.
\newblock \bibinfo{journal}{{\em interactions\/}} \bibinfo{volume}{22},
  \bibinfo{number}{2} (\bibinfo{year}{2015}), \bibinfo{pages}{24--29}.
\newblock


\bibitem[\protect\citeauthoryear{Leonard and Rayport}{Leonard and
  Rayport}{1997}]%
        {leonard1997spark}
\bibfield{author}{\bibinfo{person}{Dorothy Leonard} {and}
  \bibinfo{person}{Jeffrey~F Rayport}.} \bibinfo{year}{1997}\natexlab{}.
\newblock \showarticletitle{Spark innovation through empathic design}.
\newblock \bibinfo{journal}{{\em Harvard business review\/}}
  \bibinfo{volume}{75} (\bibinfo{year}{1997}), \bibinfo{pages}{102--115}.
\newblock


\bibitem[\protect\citeauthoryear{Lindsay, Jackson, Schofield, and
  Olivier}{Lindsay et~al\mbox{.}}{2012}]%
        {lindsay2012engaging}
\bibfield{author}{\bibinfo{person}{Stephen Lindsay}, \bibinfo{person}{Daniel
  Jackson}, \bibinfo{person}{Guy Schofield}, {and} \bibinfo{person}{Patrick
  Olivier}.} \bibinfo{year}{2012}\natexlab{}.
\newblock \showarticletitle{Engaging older people using participatory design}.
  In \bibinfo{booktitle}{{\em Proceedings of the SIGCHI conference on Human
  Factors in Computing Systems}}. ACM, \bibinfo{pages}{1199--1208}.
\newblock


\bibitem[\protect\citeauthoryear{Lum and Lightfoot}{Lum and Lightfoot}{2005}]%
        {lum2005effects}
\bibfield{author}{\bibinfo{person}{Terry~Y Lum} {and}
  \bibinfo{person}{Elizabeth Lightfoot}.} \bibinfo{year}{2005}\natexlab{}.
\newblock \showarticletitle{The effects of volunteering on the physical and
  mental health of older people}.
\newblock \bibinfo{journal}{{\em Research on aging\/}} \bibinfo{volume}{27},
  \bibinfo{number}{1} (\bibinfo{year}{2005}), \bibinfo{pages}{31--55}.
\newblock


\bibitem[\protect\citeauthoryear{Naumanen and Tukiainen}{Naumanen and
  Tukiainen}{2008}]%
        {naumanen_practices_2008}
\bibfield{author}{\bibinfo{person}{Minnamari Naumanen} {and}
  \bibinfo{person}{Markku Tukiainen}.} \bibinfo{year}{2008}\natexlab{}.
\newblock \showarticletitle{Practices in old age {ICT} education}. In
  \bibinfo{booktitle}{{\em {IADIS} {International} {Conference} on {Cognition}
  and {Exploratory} {Learning} in {Digital} {Age}}}. \bibinfo{pages}{261--269}.
\newblock


\bibitem[\protect\citeauthoryear{Nielek, Ciastek, and Kope\'{c}}{Nielek
  et~al\mbox{.}}{2017a}]%
        {nielek2017emotions}
\bibfield{author}{\bibinfo{person}{Radoslaw Nielek}, \bibinfo{person}{Miroslaw
  Ciastek}, {and} \bibinfo{person}{Wies{\l}aw Kope\'{c}}.}
  \bibinfo{year}{2017}\natexlab{a}.
\newblock \showarticletitle{Emotions Make Cities Live: Towards Mapping Emotions
  of Older Adults on Urban Space}. In \bibinfo{booktitle}{{\em Proceedings of
  the International Conference on Web Intelligence}} {\em (\bibinfo{series}{WI
  '17})}. \bibinfo{publisher}{ACM}, \bibinfo{address}{New York, NY, USA},
  \bibinfo{pages}{1076--1079}.
\newblock
\showISBNx{978-1-4503-4951-2}
\showDOI{%
\url{https://doi.org/10.1145/3106426.3109041}}


\bibitem[\protect\citeauthoryear{Nielek, Lutosta\'{n}ska, Kope\'{c}, and
  Wierzbicki}{Nielek et~al\mbox{.}}{2017b}]%
        {nielek2017wikipedia}
\bibfield{author}{\bibinfo{person}{Radoslaw Nielek}, \bibinfo{person}{Marta
  Lutosta\'{n}ska}, \bibinfo{person}{Wies{\l}aw Kope\'{c}}, {and}
  \bibinfo{person}{Adam Wierzbicki}.} \bibinfo{year}{2017}\natexlab{b}.
\newblock \showarticletitle{Turned 70?: It is Time to Start Editing Wikipedia}.
  In \bibinfo{booktitle}{{\em Proceedings of the International Conference on
  Web Intelligence}} {\em (\bibinfo{series}{WI '17})}.
  \bibinfo{publisher}{ACM}, \bibinfo{address}{New York, NY, USA},
  \bibinfo{pages}{899--906}.
\newblock
\showISBNx{978-1-4503-4951-2}
\showDOI{%
\url{https://doi.org/10.1145/3106426.3106539}}


\bibitem[\protect\citeauthoryear{Niitamo, Kulkki, Eriksson, and
  Hribernik}{Niitamo et~al\mbox{.}}{2006}]%
        {niitamo2006state}
\bibfield{author}{\bibinfo{person}{Veli-Pekka Niitamo}, \bibinfo{person}{Seija
  Kulkki}, \bibinfo{person}{Mats Eriksson}, {and} \bibinfo{person}{Karl~A
  Hribernik}.} \bibinfo{year}{2006}\natexlab{}.
\newblock \showarticletitle{State-of-the-art and good practice in the field of
  living labs}. In \bibinfo{booktitle}{{\em Technology Management Conference
  (ICE), 2006 IEEE International}}. IEEE, \bibinfo{pages}{1--8}.
\newblock


\bibitem[\protect\citeauthoryear{Orzeszek, Kopec, Wichrowski, Nielek,
  Balcerzak, Kowalik, and Puchalska-Kaminska}{Orzeszek et~al\mbox{.}}{2017}]%
        {orzeszek2017design}
\bibfield{author}{\bibinfo{person}{D. Orzeszek}, \bibinfo{person}{W. Kopec},
  \bibinfo{person}{M. Wichrowski}, \bibinfo{person}{R. Nielek},
  \bibinfo{person}{B. Balcerzak}, \bibinfo{person}{G. Kowalik}, {and}
  \bibinfo{person}{M. Puchalska-Kaminska}.} \bibinfo{year}{2017}\natexlab{}.
\newblock \showarticletitle{Beyond participatory design: Towards a model for
  teaching seniors application design}.
\newblock \bibinfo{journal}{{\em CEUR Workshop Proceed\/}}
  \bibinfo{volume}{1979} (\bibinfo{year}{2017}).
\newblock


\bibitem[\protect\citeauthoryear{Pallot, Trousse, Senach, and Scapin}{Pallot
  et~al\mbox{.}}{2010}]%
        {pallot2010living}
\bibfield{author}{\bibinfo{person}{Marc Pallot}, \bibinfo{person}{Brigitte
  Trousse}, \bibinfo{person}{Bernard Senach}, {and} \bibinfo{person}{Dominique
  Scapin}.} \bibinfo{year}{2010}\natexlab{}.
\newblock \showarticletitle{Living lab research landscape: From user centred
  design and user experience towards user cocreation}. In
  \bibinfo{booktitle}{{\em First European Summer School" Living Labs"}}.
\newblock


\bibitem[\protect\citeauthoryear{Rhode}{Rhode}{2009}]%
        {rhode2009interaction}
\bibfield{author}{\bibinfo{person}{Jason Rhode}.}
  \bibinfo{year}{2009}\natexlab{}.
\newblock \showarticletitle{Interaction equivalency in self-paced online
  learning environments: An exploration of learner preferences}.
\newblock \bibinfo{journal}{{\em The international review of research in open
  and distributed learning\/}} \bibinfo{volume}{10}, \bibinfo{number}{1}
  (\bibinfo{year}{2009}).
\newblock


\bibitem[\protect\citeauthoryear{Ries}{Ries}{2011}]%
        {ries2011lean}
\bibfield{author}{\bibinfo{person}{Eric Ries}.}
  \bibinfo{year}{2011}\natexlab{}.
\newblock \bibinfo{booktitle}{{\em The lean startup: How today's entrepreneurs
  use continuous innovation to create radically successful businesses}}.
\newblock \bibinfo{publisher}{Crown Books}.
\newblock


\bibitem[\protect\citeauthoryear{Sanders}{Sanders}{2002}]%
        {sanders2002user}
\bibfield{author}{\bibinfo{person}{Elizabeth B-N Sanders}.}
  \bibinfo{year}{2002}\natexlab{}.
\newblock \showarticletitle{From user-centered to participatory design
  approaches}.
\newblock \bibinfo{journal}{{\em Design and the social sciences: Making
  connections\/}} (\bibinfo{year}{2002}), \bibinfo{pages}{1--8}.
\newblock


\bibitem[\protect\citeauthoryear{Sanders and Stappers}{Sanders and
  Stappers}{2008}]%
        {sanders2008co}
\bibfield{author}{\bibinfo{person}{Elizabeth B-N Sanders} {and}
  \bibinfo{person}{Pieter~Jan Stappers}.} \bibinfo{year}{2008}\natexlab{}.
\newblock \showarticletitle{Co-creation and the new landscapes of design}.
\newblock \bibinfo{journal}{{\em Co-design\/}} \bibinfo{volume}{4},
  \bibinfo{number}{1} (\bibinfo{year}{2008}), \bibinfo{pages}{5--18}.
\newblock


\bibitem[\protect\citeauthoryear{Schumacher and Feurstein}{Schumacher and
  Feurstein}{2007}]%
        {schumacher2007living}
\bibfield{author}{\bibinfo{person}{Jens Schumacher} {and}
  \bibinfo{person}{Karin Feurstein}.} \bibinfo{year}{2007}\natexlab{}.
\newblock \showarticletitle{Living Labs-the user as co-creator}. In
  \bibinfo{booktitle}{{\em Technology Management Conference (ICE), 2007 IEEE
  International}}. IEEE, \bibinfo{pages}{1--6}.
\newblock


\bibitem[\protect\citeauthoryear{Szebeko and Tan}{Szebeko and Tan}{2010}]%
        {szebeko2010co}
\bibfield{author}{\bibinfo{person}{Deborah Szebeko} {and}
  \bibinfo{person}{Lauren Tan}.} \bibinfo{year}{2010}\natexlab{}.
\newblock \showarticletitle{Co-designing for Society}.
\newblock \bibinfo{journal}{{\em Australasian Medical Journal\/}}
  \bibinfo{volume}{3}, \bibinfo{number}{9} (\bibinfo{year}{2010}),
  \bibinfo{pages}{580--590}.
\newblock


\end{thebibliography}

\end{document}